\documentclass[aps,preprint,preprintnumbers,nofootinbib,showpacs]{revtex4}
\usepackage{graphicx,color,amsmath}
\begin{document}
\title{Study of $B\to \Lambda\bar\Lambda K$
and $B\to \Lambda\bar\Lambda \pi$}
\author{C. Q. Geng and Y. K. Hsiao}
\affiliation{Department of Physics, National Tsing Hua University,
Hsinchu, Taiwan 300}
\date{\today}
\begin{abstract}
We study three-body charmless baryonic $B$ decays of $B \to
\Lambda\bar\Lambda P$ with $P=\pi$ and $K$ in the standard model.
We find that the branching ratios of the $K$ modes are about one
order of magnitude larger than those of the corresponding $\pi$
modes unlike the cases of $B\to p\bar p P$. Explicitly, we obtain
that $Br(B^-\to \Lambda\bar \Lambda K^-)=(2.8\pm 0.2)\times
10^{-6}$ and $Br(\bar B^0 \to\Lambda\bar \Lambda \bar K^0)=(2.5\pm
0.3)\times 10^{-6}$. The former agrees well with the BELLE
experimental measurement of $(2.91^{+0.90}_{-0.70}\pm 0.38)\times
10^{-6}$, while the latter should be seen at the ongoing B
factories soon.

\end{abstract}
\newpage
\preprint{} \maketitle
There have been lots of attentions recently on charmless
three-body baryonic $B$ decays due to the several new experimental
measurements by BELLE and BABAR
\cite{threebody2,radiative,threebody3,moriond,Babar}. It has been
realized that baryonic $B$ decays could happen since the $B$-meson
mass can be heavier than the invariant baryon masses. In
particular, the reduced energy release can make the decays as
significant as the two-body mesonic decay modes \cite{threshold}.
As the main characteristic of the three-body baryonic decays, the
baryon pair threshold effect \cite{threshold,CK1} results in the
decay modes being accessible at the current B factories. Indeed,
some of the modes have been seen and limited recently by the BELLE
and BABAR collaborations, with the data given by
\begin{eqnarray}
 Br(B^-\to p \bar p \pi^-)&=& (3.06^{+0.73}_{-0.62}\pm 0.37)\times
 10^{-6}\;\text{(BELLE)\cite{threebody2}}\,,
 \nonumber
 \\
Br(B^0 \to p \bar p K_S)&=& (1.20^{+0.32}_{-0.22}\pm 0.14)\times
10^{-6}\;\text{(BELLE)\cite{moriond}}\,, \nonumber
\\
Br(B^- \to p\bar p K^-) &=&(5.3^{+0.45}_{-0.39}\pm 0.58)\times 10^{-6}\;\text{(BELLE)\cite{moriond}}\,,\nonumber\\
                         &&(6.7\pm 0.9\pm 0.6)\times 10^{-6}\;\text{(BABAR)\cite{Babar}}\,,\nonumber\\
Br(\bar{B}^0 \to \Lambda \bar{p}\pi^+) &=&
(3.27^{+0.62}_{-0.51}\pm 0.39)\times
10^{-6}\;\text{(BELLE)\cite{moriond}}\,,
 \nonumber
\\
Br( B^- \to \Lambda \bar p \gamma) &=& (2.16^{+0.58}_{-0.53}\pm
0.20)\times 10^{-6}\;\text{(BELLE)\cite{radiative}}\,,
 \nonumber
\\
 Br(B^- \to \Lambda\bar \Lambda K^-) &=& (2.91^{+0.90}_{-0.70}\pm 0.38)\times 10^{-6}\;\text{(BELLE)\cite{threebody3}}\,,
 \label{Data1}
  \end{eqnarray}
and
 \begin{eqnarray}
 Br(B^- \to \Lambda\bar \Lambda \pi^-)&<& 2.8\times 10^{-6}\;\text{(BELLE)\cite{threebody3}}\,,
\label{Data2}
 \end{eqnarray}
respectively. Note that the three-body decays in Eq. (\ref{Data1})
are much larger than the corresponding two-body baryonic
modes, for which
only upper bounds of $O(10^{-7})$ have been reported \cite{twobody}.

There are mainly two kinds of approaches to study the baryonic B
decays in the literature. One is the pole model, presented in
Refs. \cite{HY1,HY2,HY3}, where the intermediate particles couple
dominantly to the final states. The other, proposed in Refs.
\cite{CK1,CK2,CK3}, is based on the QCD counting rules
\cite{QCD1,QCD2}, which deal with the baryonic form factors by
power expansions. We note that global $\chi^2$ fits in Ref.
\cite{CQ} for $Br(B^- \to p \bar p \pi^-)$, $Br(B^0 \to p \bar p
K^0)$ and $Br(B^- \to p \bar p K^-)$ \cite{threebody2} with the
QCD counting rules have been performed and consistent results with
data have been derived. Furthermore,  various radiative three-body
baryonic $B$ decays \cite{CQ} have been studied.

In this report, we will concentrate on the three-body charmless baryonic decays of
\begin{eqnarray}
 B &\to&
 \Lambda\bar \Lambda
 P\ \ (P=\pi\,\ and\ K)
 \label{Decay}
 \end{eqnarray}
based on the QCD counting rules. We note that so far there has
been no theoretical study on the modes in Eq. (\ref{Decay}) in the
literature. In particular, we do not know the reason why the decay
branching ratio of $B^-\to\Lambda\bar\Lambda\pi^-$ is smaller than
that of $B^-\to\Lambda\bar\Lambda K^-$, whereas the
 corresponding $p\bar{p}$
modes are comparable, in terms of the data shown in Eqs.
(\ref{Data1}) and (\ref{Data2}). In addition, these decays are of
great interest since they provide opportunities for probing T
violating effects \cite{CQ0} due to the measurable polarization of
$\Lambda$ \cite{Suzuki}.
\begin{figure}[t!]
\centering
\includegraphics[width=2.4in]{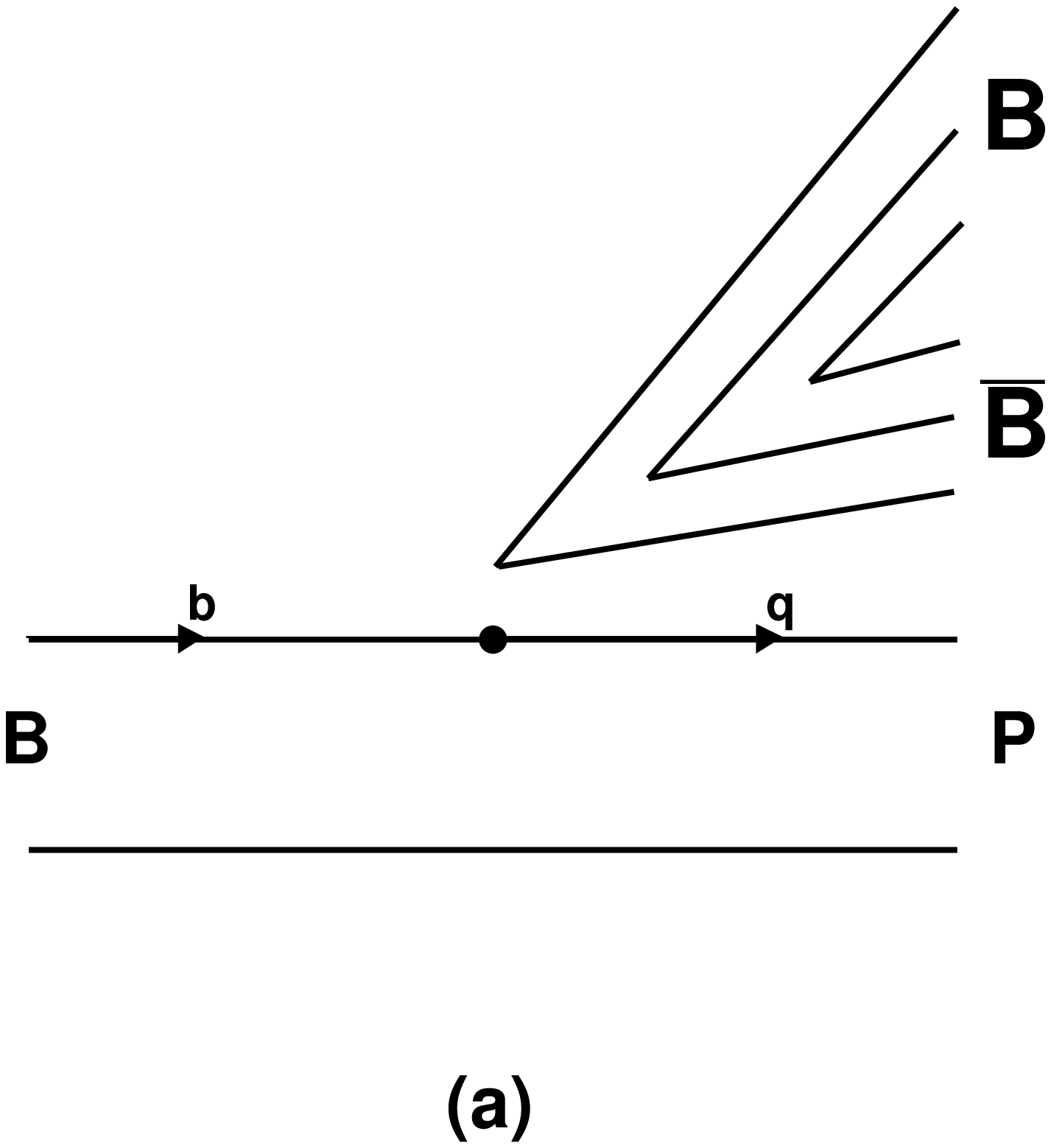}
\includegraphics[width=2.4in]{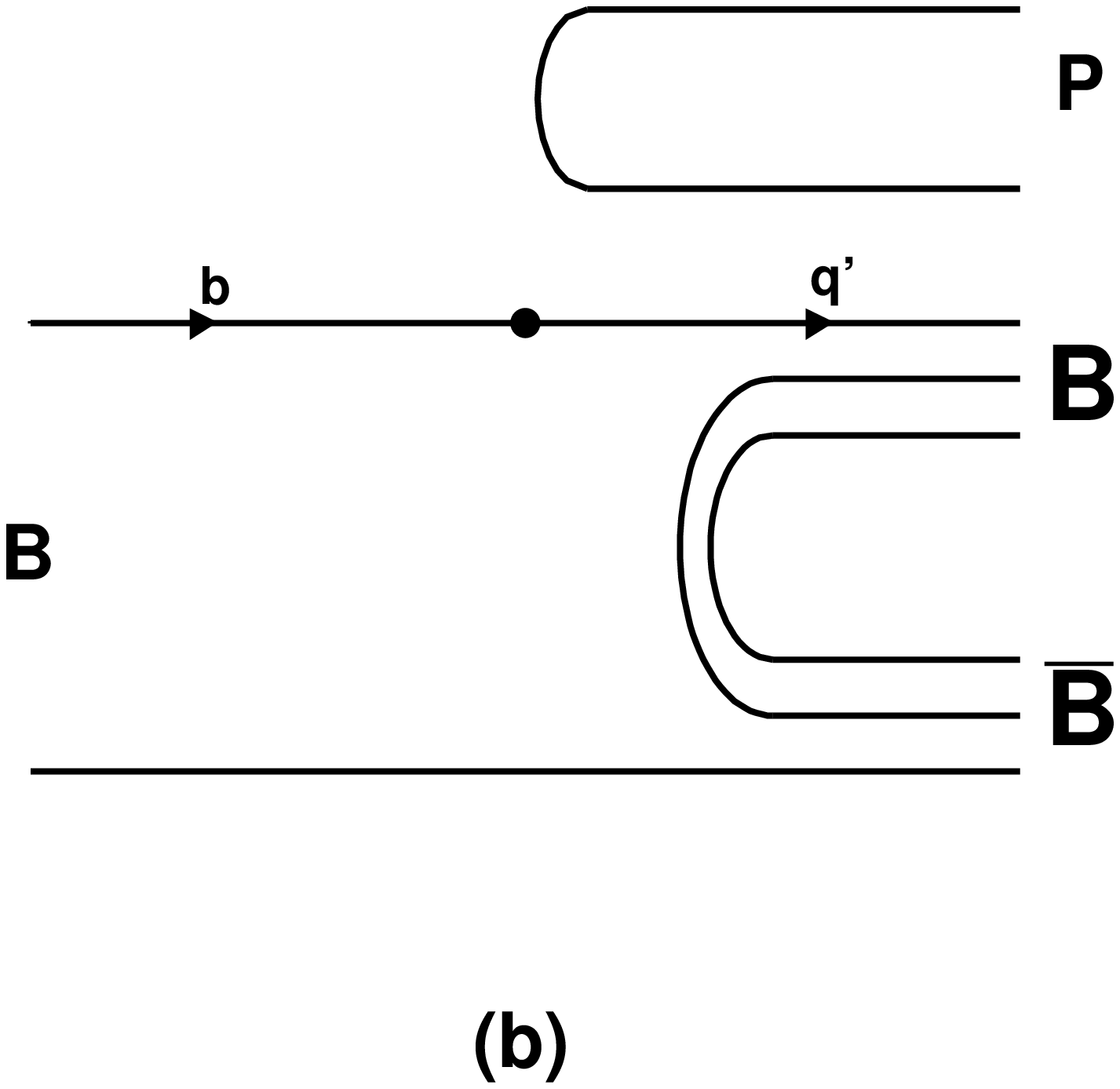}
\caption{\label{fig}
Diagrams for $B \to
{\bf B}\bar{\bf B}
P$ with ${\bf B}=p$ and $\Lambda$, $P=\pi (K)$ for $q=d(s)$ and
$q'=u(d)$ for charged (neutral) modes, and where (a) and (b)
represent ${\cal C}$ and ${\cal T}$ terms, respectively.}
\end{figure}

There are two types of diagrams which contribute to the decays in
Eq. (\ref{Decay}) under the factorization approximation
\cite{CK1},
with  Figs. \ref{fig}a and \ref{fig}b representing typical
diagrams of a current-produced baryon pair with  $B\to P$ and a
current-produced  $P$ with $B\to {\bf B}\bar{\bf B}$, named as
${\cal C}$ and ${\cal T}$ terms, respectively. The amplitude of
$B\to{\bf B}{\bf\bar B}P$ with ${\bf B}=p$ or $\Lambda$ and
$P=\pi$ or $K$ in the factorization approximation is given by
\cite{CK1}
\begin{eqnarray}
{\cal A}(B\to{\bf B}{\bf\bar B}P)={\cal C}(B\to{\bf B}{\bf\bar
B}P)+{\cal T}(B\to{\bf B}{\bf\bar B}P)\;,
\end{eqnarray}
where \cite{ali,cheng}
\begin{eqnarray} {\cal C}(B\to{\bf
B}{\bf\bar B}P)&=&\frac{G_F}{\sqrt 2}\bigg\{ V_{ub}V_{uq}^*
a_2\langle {\bf B}{\bf\bar B}|(\bar u u)_{V-A}|0\rangle
\langle P|(\bar q b)_{V-A}|B\rangle\nonumber\\
&&-V_{tb}V_{tq}^*\bigg[a_3\langle {\bf B}{\bf\bar B}|(\bar u u+\bar d d+\bar s s)_{V-A}|0\rangle
+a_4\langle {\bf B}{\bf\bar B}|(\bar q q)_{V-A}|0\rangle\nonumber\\
&&
+a_5\langle {\bf B}{\bf\bar B}|(\bar u u+\bar d d+\bar s s)_{V+A}|0\rangle
+\frac{3}{2}a_9\langle {\bf B}{\bf\bar B}|(e_u\bar u u+e_d\bar d
d+e_s\bar s s)_{V-A}|0\rangle\bigg] \nonumber\\&&
\langle P|(\bar q b)_{V-A}|B\rangle
+V_{tb}V_{tq}^*2a_6\langle {\bf B}{\bf\bar B}|(\bar q
q)_{S+P}|0\rangle\langle P |(\bar q b)_{S-P}|B\rangle\bigg\}\;,
\label{Jterm}
\end{eqnarray}
with $q=d(s)$ for $P=\pi(K)$, while ${\cal T}(B\to{\bf B}{\bf\bar
B}P)$ is decomposed with the $P$-meson decay constant $f_P$
induced from the $P$ creation and $B \to {\bf B}{\bf \bar B}$
through scalar and pseudoscalar currents as
\begin{eqnarray}\label{tterm1}
{\cal T}(B\to{\bf B}{\bf\bar B}P)&=&i\frac{G_F}{\sqrt
2}f_{P}m_b\bigg[\alpha_{P}\langle {\bf B}{\bf\bar B}|\bar
q'b|B\rangle +\beta_{P}\langle {\bf B}{\bf\bar B}|\bar q'\gamma_5
b|B\rangle\bigg]\;,
\end{eqnarray}
with $q'=u(d)$ for charged (neutral) modes, and
\begin{eqnarray}\label{tterm11}
\alpha_{P}(\beta_{P}) &=&
\bigg[V_{ub}V_{uq}^*a_P-V_{tb}V_{tq}^*\bigg(a_P'\pm
a_P''\frac{2m_{P}^2}{m_b(m_q+m_{q'})}\bigg)\bigg]\,,
\end{eqnarray}
where $a_{\pi^-}=a_{K^-}=a_1$, $a_{\pi^0}=a_2$, $a_{K^0}=0$,
$a_{\pi^-}'=a_{K^-}'=a_{K^0}'=a_4$ ,$a_{\pi^0}'=-a_4+3a_9/2$,
$a_{\pi^-,K^-,K^0}''=a_6$ and $a_{\pi^0}''=-a_6$. Here,
$f_{\pi^0}=f_{\pi}/\sqrt 2$, $V_{ij}$ denote the CKM mixing matrix
elements
 \cite{CKM}, $a_i$ ($i=1,\cdots, 10$)  are composed with (effective) Wilson coefficients
$c_i^{(eff)}\;(i=1,2,\cdots,10)$,
 defined in
Refs. \cite{ali,cheng}, and $N_c$ is the effective color number.
The coefficients $a_i$ and $c_i^{(eff)}$ are related by
\begin{eqnarray}
a_{i}=c_{i}^{eff}+\frac{c_{i+1}^{eff}}{N_c}\;\;(\text{$i$=odd})\;,\;\;\;a_i=c_i^{eff}+\frac{c_{i-1}^{eff}}{N_c}\;\;(\text{$i$=even})\;.
\end{eqnarray}

To calculate the decay rate, we need to know the hadronic transition matrix elements in Eqs. (\ref{Jterm}) and (\ref{tterm1}).
 The $ B\to P$ transition matrix element can be parameterized as
\begin{eqnarray}\label{form}
\langle P| \bar q \gamma^\mu(1-\gamma_5)b|B\rangle&=&
\bigg[(p_B+p_P)^\mu-\frac{m^2_B-m^2_P}{(p_B-p_P)^2}(p_B-p_P)^\mu\bigg
]F_1^{B\to P}(t)\nonumber\\
&+&\frac{m^2_B-m^2_P}{(p_B-p_P)^2}(p_B-p_P)^\mu F_0^{B\to P}(t)\,,
\end{eqnarray}
where $t\equiv (p_{\bf B}+p_{\bf\bar B})^2$ and $F_{1,0}^{B\to
P}(t)$ are defined by \cite{MS}
\begin{eqnarray}\label{form2}
F^{B \to P}_1(t)&=&\frac{F^{B \to P}_1(0)}{(1-\frac{t}{M_V^2})(1-\frac{\sigma_{11} t}{M_V^2}+\frac{\sigma_{12} t^2}{M_V^4})}\;,\nonumber\\
F^{B \to P}_0(t)&=&\frac{F^{B \to P}_0(0)}{1-\frac{\sigma_{01}
t}{M_V^2}+\frac{\sigma_{02} t^2}{M_V^4}}\;,
\end{eqnarray}
with the input parameter values of $F^{B \to \pi}_1(0)=F^{B \to
\pi}_0(0)=0.29$, $\sigma_{11}=0.48$, $\sigma_{12}=0$,
$\sigma_{01}=0.76$, $\sigma_{02}=0.28$ and $M_V=5.32$ GeV for $B
\to \pi$ and
 $F^{B \to K}_1(0)=F^{B
\to K}_0(0)=0.36$, $\sigma_{11}=0.43$, $\sigma_{12}=0$,
$\sigma_{01}=0.70$, $\sigma_{02}=0.27$ and $M_V=5.42$ GeV for $B
\to K$, respectively. For those of the baryon pair involving the
vector, axial-vector, scalar and pseudoscalar currents in Eq.
(\ref{Jterm}), we have
\begin{eqnarray}\label{form3}
\langle {\bf B}{\bf\bar B}|V_{\mu}|0\rangle &=&\bar
u(p_{\bf B})\bigg\{F_1(t)\gamma_\mu+\frac{F_2(t)}{m_{\bf B}+m_{\bf \bar B}}
\sigma_{\mu\nu}(p_{\bf \bar B}+p_{\bf B})_\mu\bigg\}v(p_{\bf \bar B})\;,\nonumber\\
&=& \bar
u(p_{\bf B})\bigg\{[F_1(t)+F_2(t)]\gamma_\mu+\frac{F_2(t)}{m_{\bf B}+m_{\bf \bar B}}
(p_{\bf \bar B}-p_{\bf B})_\mu\bigg\}v(p_{\bf \bar B})\;,\nonumber\\
\langle {\bf B}{\bf\bar B}|A_\mu|0\rangle&=&\bar
u(p_{\bf B})\bigg\{g_A(t)\gamma_\mu+\frac{h_A(t)}{m_{\bf B}+m_{\bf \bar B}}
(p_{\bf \bar B}+p_{\bf B})_\mu\bigg\}\gamma_5 v(p_{\bf \bar B})\,,
\nonumber\\
\langle {\bf B}{\bf\bar B}|S|0\rangle &=&f_S(t)\bar
u(p_{\bf B})v(p_{\bf\bar B})\;,\nonumber\\
\langle {\bf B}{\bf\bar B}|P|0\rangle &=&g_P (t)\bar u(p_{\bf
B})\gamma_5 v(p_{\bf\bar B})\,,
\end{eqnarray}
where $V_\mu=\bar q_i\gamma_\mu q_j$, $A_\mu=\bar
q_i\gamma_\mu\gamma_5 q_j$, $S=\bar q_i q_j$ and $P=\bar
q_i\gamma_5 q_j$ with $q_i=u,d$ and $s$. We note that $F_2(t)$
alone can not be determined by the present experimental data.
However, $F_2(t)$ can be ignored since it acquires one more $1/t$
than $F_1(t)$ according to the power expansion in a perturbative
QCD re-analysis \cite{reanalysis1,reanalysis2}. By using  equation
of motion and adopting  zero quark mass limits, we have
\begin{eqnarray}
h_A(t)= -g_A(t)\frac{(m_{\bf B}+m_{\bf\bar B})^2}{t}.
\end{eqnarray}
In terms of the $SU(3)$ flavor symmetry,
 the form factors $[F_1(t)+F_2(t)]$, $g_A(t)$, $f_S(t)$
and $g_P(t)$  in Eq. (\ref{form3})  can be related to another set
of form factors $D_X(t)$, $F_X(t)$ and $S_X(t)$ where $X=V,A,S$
and $P$
 denote vector, axial-vector, scalar, pseudo-scalar
currents, defined in Table \ref{table1}, respectively. It is noted
that the zero value of $\langle p\bar p|(\bar ss)_X|0\rangle$ in
the second column is due to the OZI suppression rule.
\begin{table}[h!]
\caption{ Relations of form factors between different sets of
parameterizations, where the form factors $[F_1(t)+F_2(t)]$,
$g_A(t)$, $f_S(t)$ and $g_P(t)$ correspond to the notations $V$,
$A$, $S$ and $P$ in the first column, respectively.}\label{table1}
\begin{tabular}{|c||c|c|c|}
\hline Form Factor (X=V,A,S,P)&${\bf B}{\bf\bar B}=p\bar p$&${\bf B}{\bf\bar B}=\Lambda\bar \Lambda$\\\hline
$\langle {\bf B}{\bf\bar B}|(\bar u u)_{X}|0\rangle$&$D_X(t)+F_X(t)+S_X(t)$&$\frac{1}{3}[D_X(t)+3S_X(t)]$\\
$\langle {\bf B}{\bf\bar B}|(\bar d d)_{X}|0\rangle$&$S_X(t)$&$\frac{1}{3}[D_X(t)+3S_X(t)]$\\
$\langle {\bf B}{\bf\bar B}|(\bar s s)_{X}|0\rangle$&$D_X(t)-F_X(t)+S_X(t)=0$&$\frac{1}{3}[4D_X(t)+3S_X(t)]$\\
$\langle {\bf B}{\bf\bar B}|(\bar u u+\bar d d+\bar s s)_{X}|0\rangle$&$2D_X(t)+3S_X(t)$&$2D_X(t)+3S_X(t)$\\
$\langle {\bf B}{\bf\bar B}|(e_u\bar u u+e_d\bar d d+e_s\bar s s)_{X}|0\rangle$&$F_X(t)+\frac{1}{3}D_X(t)$ &$-\frac{1}{3}D_X(t)$\\
\hline
\end{tabular}
\end{table}
\normalsize
As an illustration,
we take $\langle \Lambda\bar \Lambda|(\bar u u+\bar d
d+\bar s s)_{V,A}|0\rangle$ in Eq. (\ref{Jterm}) and we have
\begin{eqnarray}
F_1^{\Lambda\bar \Lambda}(t)+F_2^{\Lambda\bar \Lambda}(t)&=&2D_V(t)+3S_V(t)\;,\nonumber\\
g_A^{\Lambda\bar \Lambda}(t)&=&2D_A(t)+3S_A(t)\;,\nonumber\\
h_A^{\Lambda\bar \Lambda}(t)&=&-g_A^{\Lambda\bar
\Lambda}(t)\frac{(m_{\Lambda}+m_{\bar \Lambda})^2}{t}\;.
\end{eqnarray}
For
$D_X(t)$, $F_X(t)$, and $S_X(t)$, since they are related to the
nucleon magnetic (Sachs) form factors $G^{p(n)}_M(t)$, we adopt
the results for  $G^{p(n)}_M(t)$ in Ref. \cite{CK4}, which are
extracted from experiments.

  The functions of $G^{p(n)}_M(t)$,
$D_X(t)$, $F_X(t)$ and $S_X(t)$ are parameterized as
\begin{eqnarray}\label{expand}
G^p_M(t)=\displaystyle\sum^{5}_{i=1}\frac{x_i}{t^{i+1}}\left[\text{ln}(\frac{t}{\Lambda^2_0})\right]^{-\gamma},
G^n_M(t)=\displaystyle\sum^2_{i=1}\frac{y_i}{t^{i+1}}\big[\text{ln}(\frac{t}{\Lambda^2_0})\big]^{-\gamma},
S_V(t)=\displaystyle\sum^2_{i=1}\frac{s_i}{t^{i+1}}\big[\text{ln}(\frac{t}{\Lambda^2_0})\big]^{-\gamma},\nonumber\\
D_A(t)=\displaystyle\sum^2_{i=1}\frac{\tilde{d}_i}{t^{i+1}}\left[\text{ln}(\frac{t}{\Lambda^2_0})\right]^{-\gamma},
F_A(t)=\displaystyle\sum^2_{i=1}\frac{\tilde{f}_i}{t^{i+1}}\left[\text{ln}(\frac{t}{\Lambda^2_0})\right]^{-\gamma},
S_A(t)=\displaystyle\sum^2_{i=1}\frac{\tilde{s}_i}{t^{i+1}}\left[\text{ln}(\frac{t}{\Lambda^2_0})\right]^{-\gamma},\nonumber\\
D_P(t)=\displaystyle\sum^2_{i=1}\frac{\bar d_i}{t^{i+1}}\left[\text{ln}(\frac{t}{\Lambda^2_0})\right]^{-\gamma},
F_P(t)=\displaystyle\sum^2_{i=1}\frac{\bar f_i}{t^{i+1}}\left[\text{ln}(\frac{t}{\Lambda^2_0})\right]^{-\gamma},
S_P(t)=\displaystyle\sum^2_{i=1}\frac{\bar s_i}{t^{i+1}}\left[\text{ln}(\frac{t}{\Lambda^2_0})\right]^{-\gamma},\nonumber\\
\end{eqnarray}
with
\begin{eqnarray}
 D_V(t)&=&-{3\over 2}G^n_M(t)\,,\ F_V(t)=G^p_M(t)+{1\over 2}G^n_M(t)\,,\
D_S(t)=-{3\over 2}n_qG^n_M(t)\,,\nonumber\\
F_S(t)&=&n_q[G^p_M(t)+{1\over 2}G^n_M(t)]\,, \
S_S(t)=n_q S_V(t)\,.
\label{nq}
\end{eqnarray}
The input numbers can be found in Refs. \cite{CK1,CK2,CK3}.
Explicitly, we take $\gamma=2.148$, $x_1=420.96\;\text{GeV}^4$,
$x_2=-10485.50\;\text{GeV}^6$, $x_3=106390.97\;\text{GeV}^8$,
$x_4=-433916.61\;\text{GeV}^{10}$,
$x_5=613780.15\;\text{GeV}^{12}$, $y_1=292.62\;\text{GeV}^4$,
$y_2=-579.51\;\text{GeV}^6$, $s_1=x_1-2y_1$,
$s_2=500\;\text{GeV}^6$, $\tilde{d}_1=x_1-\frac{3}{2}y_1$,
$\tilde{f}_1=\frac{2}{3}x_1+\frac{1}{2}y_1$,
$\tilde{s}_1=-x_1/3+2y_1$,
$\tilde{d}_2=\tilde{f}_2=-478\;\text{GeV}^6$, $\tilde{s_2}=0$,
$\bar d_1=n_q \frac{3}{2}y_1$, $\bar f_1=n_q(x_1-\frac{3}{2}y_1)$,
$\bar s_1=n_q(x_1-2y_1)$, $\bar s_2=n_q s_2$, $\bar d_2=\bar
f_2=-952\;\text{GeV}^6$ and $\Lambda_0=0.3\;\text{GeV}$. We remark
that the parameter $n_q$ in Eq. (\ref{nq}) corresponds to the
$(m_{\bf B}-m_{\bf \bar B'})/(m_q-m_{\bar q'})$ term in connecting
scalar form factors to vector ones in the case of ${\bf B}\neq{\bf
\bar B'}$. In $B\to {\bf B\bar B}P$ decays, this parameter is not
well-defined. However, by taking $m_{\bar q'}\rightarrow m_{q}$
and $m_{\bf \bar B'}\rightarrow m_{\bf B}$, it has been shown
\cite{CK3} that $n_q$ is around $1.3-1.4$. Here we fix $n_q\simeq
1.4$.

For the $B\to{\bf B}{\bf\bar B}$ transition in Eq. (\ref{tterm1}),
we first discuss the case with ${\bf B}=p$. For $B^-\to p\bar p$,
one has \cite{CK1}
\begin{eqnarray}\label{pp1}
\langle p \bar p|\bar u b|B^-\rangle&=&i\bar u(p_{p})
[F_A \not{\!p}\gamma_5+F_P \gamma_5]v(p_{\bar p})\;,\nonumber\\
\langle p \bar p|\bar u\gamma_5 b|B^-\rangle&=&i\bar u(p_{p})
[F_V \not{\!p}+F_S]v(p_{\bar p})\,,
\end{eqnarray}
where $p=p_B-(p_{\bar p}+p_p)$. Note that $F_S=F_P$ as shown in
Ref. \cite{CK3}.
For $\bar B^0\to p\bar p$, one gets
\begin{eqnarray}\label{pp2}
\langle p \bar p|\bar d b|\bar B^0\rangle&=&i\bar u(p_{p})
[F_A^{p \bar p} \not{\!p}\gamma_5+F_P^{p \bar p} \gamma_5]v(p_{\bar p})\;,\nonumber\\
\langle p \bar p|\bar d\gamma_5 b|\bar B^0\rangle&=&i\bar u(p_{p})
[F_V^{p \bar p} \not{\!p}+F_S^{p \bar p}]v(p_{\bar p})\;.
\end{eqnarray}
The form factors in Eqs. (\ref{pp1}) and (\ref{pp2}) can be
related \cite{CK1,CQ} by the SU(3) symmetry and the helicity
conservation \cite{QCD1,QCD2} and one obtains that
\begin{eqnarray}\label{form1}
F^{p \bar p}_{A}=\frac{1}{10}(11F_{A}+9F_V),\;\;F^{p \bar
p}_{V}=\frac{1}{10}(9F_A+11F_V),\;\; F^{p \bar p}_{P(S)}=-\frac{1 }{4}F_P.
\end{eqnarray}
Moreover, the three form factors $F_A$, $F_V$ and $F_P$
in Eq. (\ref{pp1})
can be simply
presented by \cite{CK1}
\begin{eqnarray}\label{para}
F_{A,V}=\frac{C_{A,V}}{t^3},\;\;\;F_{P}=\frac{C_P}{t^4}\,,
\end{eqnarray}
where $C_i\; (i=A,V,P)$ are new parameterized form factors, which
are taken to be real. By following the approach of Refs.
\cite{CK1,QCD2}, the form factors in $B^{-,0} \to \Lambda\bar
\Lambda$ transitions are given by
\begin{eqnarray}\label{form2}
F^{\Lambda\bar\Lambda}_{A}=\frac{C_{A}^{\Lambda\bar\Lambda}}{t^3},\;\;
F^{\Lambda\bar\Lambda}_{V}=\frac{C_{V}^{\Lambda\bar\Lambda}}{t^3},\;\;
F^{\Lambda\bar\Lambda}_{P(S)}=\frac{C_{P}^{\Lambda\bar\Lambda}}{t^4}\;,
\end{eqnarray}
where
\begin{eqnarray}\label{form22}
C_{A}^{\Lambda\bar\Lambda}=\frac{1}{10}(9C_{A}+6C_V)\;,\;\;
C_{V}^{\Lambda\bar\Lambda}=\frac{1}{10}(6C_{A}+9C_V)\;,\;\;
C_{P}^{\Lambda\bar\Lambda }=0.
\end{eqnarray}
It is interesting to note that $F^{\Lambda\bar \Lambda}_{P(S)}=0$
from Eqs. (\ref{form2}) and (\ref{form22}). The reason\footnote{We
thank the argument given by the referee.} for this result is that
$F^{\Lambda\bar \Lambda}_{P(S)}$ correspond to a helicity
(chirality) flipped terms in the $B \to \Lambda\bar\Lambda$ matrix
elements in the large momentum transfer. It is well known that the
spin of $\Lambda$ is carried by the s-quark component. Hence, such
a term requires a chirality flip in the s-quark, which is absent
in the decay amplitude of $B\to\Lambda\bar\Lambda$.

To obtain these unknown form factors, we use the $\chi^2$ fit with
the experimental data in Eq. (\ref{Data1}).
 Here we have neglected \cite{CQ} the $C_P$ term since it
has one more 1/t over $C_A$ and $C_V$ as shown in Eq.
(\ref{para}). Therefore, we keep the numbers of the degree of
freedom (ndf) to be 2. In our fit, we also include the
uncertainties from the Wolfenstein parameters \cite{wolfenstein}
in the CKM matrix. Explicitly, we use $\lambda=0.2200\pm 0.0026$,
$A=|V_{cb}|/\lambda^2=0.853\pm0.037$, $\rho=0.20\pm 0.09$ and
$\eta=0.33\pm 0.05$ \cite{pdg}. We follow the Refs.
\cite{cheng,cheng2} to deal with $a_i$ ($i=1, ...,10$) and we take
Wilson coefficients from Ref. \cite{WC}.

The experimental inputs and the fitted results are shown in Table
\ref{parameter}.
\begin{table}[t!]
\begin{center}
\caption{Fits of ($C_A$,$C_V$) in units of $GeV^4$.
 }\label{parameter}
\begin{tabular}{|c|c||c|c|}
\hline Input&experimental data&Fit result&best fit (with 1$\sigma$
error )\\\hline $Br(B^- \to p \bar p \pi^-)$ \cite{threebody2}&
Eq. (\ref{Data1}) &$C_A$&$-71.0\pm 5.3$\\
$Br(B^0 \to p \bar p K^0)$ \cite{moriond}&
&$C_V$&\,\,$42.0\pm 9.2$\\
$Br(B^- \to p \bar p K^-)$\cite{moriond}&
&$\chi^2/ndf$&1.9\\
$Br(B^- \to \Lambda \bar p \gamma)$ \cite{radiative}&
&&\\\hline
\end{tabular}
\end{center}
\end{table}
\normalsize With the fitted values in Table \ref{parameter}, the
theoretical predictions with 1$\sigma$ error for
$B^-\to\Lambda\bar{\Lambda}P$ are given by
\begin{eqnarray}
Br(B^- \to \Lambda \bar \Lambda K^-)&=&(2.8\pm 0.2)\times 10^{-6}\;,\nonumber\\
Br(B^- \to \Lambda \bar \Lambda \pi^-)&=&(1.7\pm 0.7)\times10^{-7}\;, \label{BR1}
\end{eqnarray}
which agree very well with the recent BELLE data \cite{threebody3}
in Eq. (\ref{Data1}). We note that the spectrum of
$B^-\to\Lambda\bar\Lambda K^-$ shown in Fig. \ref{fig}b is
consistent with that in Ref. \cite{threebody3} by BELLE. Our
predicted values in Eq. (\ref{BR1}) show that $Br(B^-\to
\Lambda\bar \Lambda\pi^-)\ll Br(B^-\to \Lambda\bar \Lambda K^-)$
unlike $Br(B^- \to p \bar p \pi^-)\sim Br(B^-\to p\bar p K^-)$ as
indicated in Eq. (\ref{Data1}). This result can be easily
understood from the theoretical point of view. As $B^- \to p \bar
p \pi^-$ can not obtain a large contribution from $\cal{C}$ in Eq.
(\ref{Jterm}) since $a_2$ is color suppressed, its main
contribution is from $a_1$ term in $\cal{T}$  as seen in Eqs.
(\ref{tterm1}) and (\ref{tterm11}), whereas $B^-\to p\bar p K^-$
is due to the penguin part which gives contributions to the terms
in both $\cal{C}$ and $\cal{T}$. However, it is not the case for
$B^-\to \Lambda\bar \Lambda K^-$ since $\langle \Lambda\bar
\Lambda|(\bar ss)_X|0\rangle$ in Eq. (\ref{Jterm}) escapes from
the OZI suppression. Its contribution is mainly from $\cal C$
which is enhanced by $a_6$ with the chiral enhancement, whereas it
is suppressed in $\cal{T}$. For $B^- \to \Lambda\bar \Lambda
\pi^-$, on the other hand, because of the $a_2$ color suppression
in $\cal{C}$ and the small contribution in $\cal{T}$, its
branching ratio is small.
\begin{figure}[t!]
\centering
\includegraphics[width=3.0in]{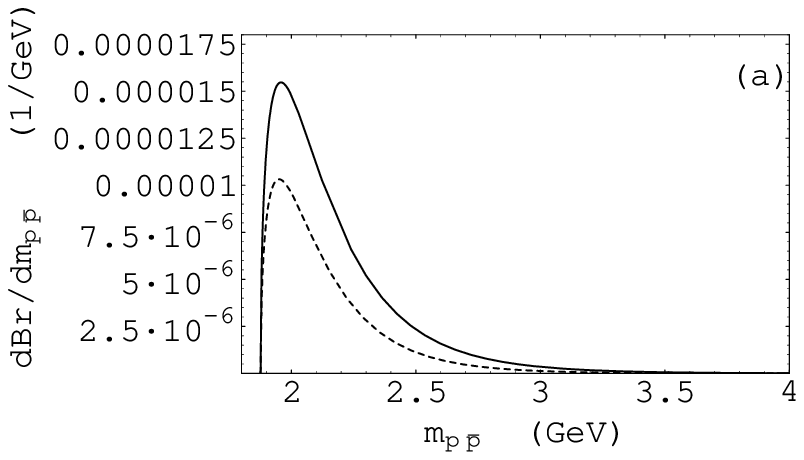}
\includegraphics[width=3.0in]{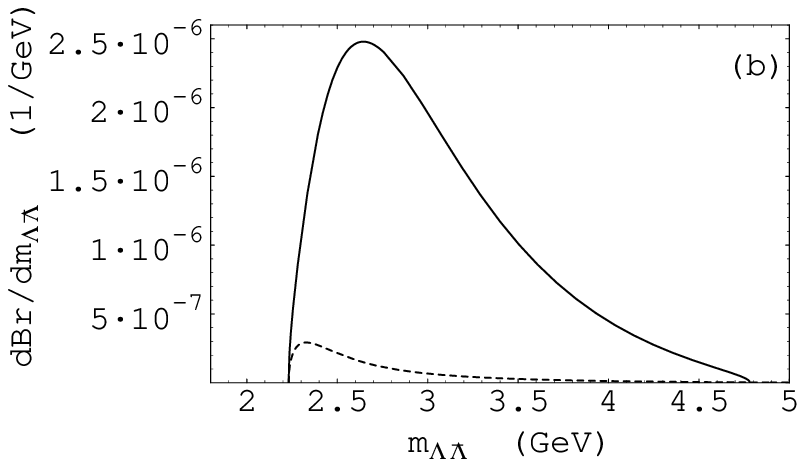}
\caption{$dBr$/$dm_{\bf B\bar B}$ as a function of $m_{\bf B\bar
B}$ for (a) $B^-\to p\bar p P$ and (b) $B^-\to\Lambda\bar\Lambda
P$, where the solid and dashed lines for $P=K^-$ and $P=\pi^-$,
respectively.}\label{fig}
\end{figure}
Moreover, to explicitly see the $\cal{T}$-suppression in the
$\Lambda\bar\Lambda$ case, we show the spectra of $B^- \to p\bar
p\pi^-$ and $B^-\to\Lambda\bar\Lambda\pi^-$ in Fig. \ref{fig}. We
note that the two modes contain the same $\cal{T}$-amplitudes in
Eq. (\ref{tterm1}). As shown in Fig. \ref{fig} the main
contributions to the rates are due to the threshold effect,
resulting from the form factors of $F_{A,V}={C_{A,V}}/{t^3}$ and
$F_{A,V}^{\Lambda\bar\Lambda}={C_{A,V}^{\Lambda\bar\Lambda}}/{t^3}$
in $B^- \to p\bar p\pi^-$ and $B^-\to\Lambda\bar\Lambda\pi^-$,
respectively. When $t=(p_{\bf B}+p_{\bf\bar B})^2\rightarrow
t_{min}=(m_{\bf B}+m_{\bf \bar B})^2$,
$t^{-3}_{\Lambda\bar\Lambda}\simeq t^{-3}_{p\bar p}/3$.
Furthermore, the relations from the numerator of the form factor
are ${C_{A}^{\Lambda\bar\Lambda}}\simeq {C_{A}}/2$ and
${C_{V}^{\Lambda\bar\Lambda}}\simeq -{C_{V}}/10$. Once we combine
the relations above and integrate them through the phase space, we
obtain that $Br(B^-\to\Lambda\bar\Lambda\pi^-)\simeq
O(10^{-1})Br(B^- \to p\bar p\pi^-)$.

Similarly, we can also study the neutral decay modes of $\bar B^0
\to \Lambda \bar \Lambda \bar K^0(\pi^0 )$, which have not been
measured yet. The decay branching ratios are found to be
\begin{eqnarray}
Br(\bar B^0\to{\Lambda}{\bar \Lambda} \bar K^0)&=&(2.5\pm 0.3)\times
10^{-6},\nonumber\\
Br(\bar B^0\to{\Lambda}{\bar \Lambda} \pi^0)&=&(0.4\pm 0.4)\times
10^{-7}.
\label{BR2}
\end{eqnarray}
As seen from Eq. (\ref{BR2}), the decay branching ratio of $\bar
B^0\to{\Lambda}{\bar \Lambda} \bar K^0$ is almost the same as that
of $ B^- \to \Lambda \bar \Lambda K^-$, whereas $\bar
B^0\to{\Lambda}{\bar \Lambda} \pi^0$ is still suppressed as
$B^-\to{\Lambda}{\bar \Lambda} \pi^-$.
We note that
the errors of
$Br(B^-\to{\Lambda}{\bar \Lambda} K^-)$ and $Br(\bar
B^0\to{\Lambda}{\bar \Lambda} \bar K^0)$
in Eqs. (\ref{BR1}) and (\ref{BR2})
are small since the
main contributions are not from $C^{\Lambda\bar \Lambda}_{A,V}$
which receive almost all uncertainties from the data.

To describe the
possible non-factorizable effects, we also fit the data with
$N_c=2$ and $\infty$.
 As expected, we find that
the branching ratios of the $K$ modes are  slightly changed, while
the central values of $B^- \to \Lambda\bar\Lambda\pi^-$ and $\bar
B^0 \to \Lambda\bar\Lambda\pi^0$ shift to 2.3  and 1.3 (0.7  and
0.1) for $N_c=2~(\infty)$, respectively. We may conclude that the
two $\pi$ modes remain small even with including all possible
non-factorizable effects. However, as pointed in Ref. \cite{a2},
$a_2$ can only be determined by the experimental data in the
two-body $B$ decays, since the experimental data of $Br(\bar
B^0\to \pi^0\pi^0)$ and $Br(\bar B^0\to D^0\pi^0)$ are much larger
than the theoretical values, which means the failure of the
factorization approximation. On the other hand, in the three-body
baryonic $B$ decays, due to the complicated topology of Feynman
diagrams, it is not as easy as those of the two-body decays to
influence $a_2$ by the annihilations \cite{anni} as well as the
final state interactions \cite{FSI}. Therefore, the value of $a_2$
may not change much. Nevertheless, we leave the surprise to the
experimentalists if the factorization does not work well in these
two modes as those in the two-body decays.

In sum, we have studied three-body charmless baryonic $B$ decays
of $B \to \Lambda\bar\Lambda \pi$ and $B \to \Lambda\bar\Lambda K$
based on the QCD counting rules in the standard model. We have
shown that $Br(B^{-,0} \to \Lambda\bar\Lambda K^{-,0})\gg
Br(B^{-,0} \to \Lambda\bar\Lambda \pi^{-,0})$ unlike the cases of
$B\to p\bar p \pi\ (K)$. Explicitly, we have found that $Br(B^-\to
\Lambda\bar \Lambda K^-)=(2.8\pm 0.2)\times 10^{-6}$, $Br(\bar B^0
\to\Lambda\bar \Lambda \bar K^0)=(2.5\pm 0.3)\times 10^{-6}$,
$Br(B^-\to \Lambda\bar \Lambda \pi^-)=(1.7\pm 0.7)\times 10^{-7}$
and $Br(\bar{B}^0\to \Lambda\bar \Lambda \pi^0)=(0.4\pm 0.4)\times
10^{-7}$. It is interesting to note that the decay of $\bar B^0
\to\Lambda\bar \Lambda \bar K^0$ should be seen at the ongoing B
factories soon.
\section*{Acknowledgements}
This work was supported in part by
 the National Science Council of the Republic of China under
 Contract \#s:
 No. NSC-93-2112-M-007-014 and NSC-93-2112-M-007-025.

\end{document}